\documentclass[aps,showkeys,twocolumn,showpacs,preprintnumbers,amsmath,amssymb,superscriptaddress,floatfix,nofootinbib]{revtex4}
\usepackage{graphicx,color,dcolumn,booktabs,bm}
\usepackage{longtable,lscape}
\usepackage{txfonts}
\usepackage{overpic}
\usepackage{epsfig}
\usepackage{amssymb}
\usepackage{rotating}
\usepackage{epstopdf}
\usepackage{indentfirst}
\usepackage{feynmf}   %{feynmp}
\usepackage{slashed}  %for Feynman symbols
\usepackage{cases}
\usepackage{color}
\usepackage{multirow}
\usepackage{graphicx,color,dcolumn,booktabs,bm}
\usepackage{cases}
\usepackage{array}

\begin{document}

\title{Strong decay of $\Lambda_c(2940)$ as a $2P$ state in the $\Lambda_c$ family }

\author{Qi-Fang L\"{u}} \email{lvqifang@hunnu.edu.cn}
\affiliation{Department of Physics, Hunan Normal University, and Key Laboratory of Low-Dimensional Quantum
Structures and Quantum Control of Ministry of Education, Changsha 410081, China}
\affiliation{Synergetic Innovation Center for Quantum Effects and Applications (SICQEA),
Hunan Normal University, Changsha 410081, China}
\author{Li-Ye Xiao} \email{lyxiao@pku.edu.cn}
\affiliation{School of Physics and State Key Laboratory of Nuclear Physics and Technology, Peking University, Beijing 100871, China}
\affiliation{Center of High Energy Physics, Peking University, Beijing 100871, China}
\author{Zuo-Yun Wang}
\affiliation{Department of Physics, Hunan Normal University, and Key Laboratory of Low-Dimensional Quantum
Structures and Quantum Control of Ministry of Education, Changsha 410081, China}
\affiliation{Synergetic Innovation Center for Quantum Effects and Applications (SICQEA),
Hunan Normal University, Changsha 410081, China}
\author{Xian-Hui Zhong} \email{zhongxh@hunnu.edu.cn}
\affiliation{Department of Physics, Hunan Normal University, and Key Laboratory of Low-Dimensional Quantum
Structures and Quantum Control of Ministry of Education, Changsha 410081, China}
\affiliation{Synergetic Innovation Center for Quantum Effects and Applications (SICQEA),
Hunan Normal University, Changsha 410081, China}

\begin{abstract}

Considering the mass, parity and $D^0 p$ decay mode, we tentatively assign the $\Lambda_c(2940)$ as the $P-$wave states with one radial excitation. Then, via studying the strong decay behavior of the $\Lambda_c(2940)$ within the $^3P_0$ model, we obtain that the total decay widths of the $\Lambda_{c1}(\frac{1}{2}^-,2P)$ and $\Lambda_{c1}(\frac{3}{2}^-,2P)$ states are 16.27 MeV and 25.39 MeV, respectively. Compared with the experimental total width $27.7^{+8.2}_{-6.0}\pm0.9^{+5.2}_{-10.4}~\rm{MeV}$ measured by LHCb Collaboration, both assignments are allowed, and the $J^P=\frac{3}{2}^-$ assignment is more favorable. Other $\lambda-$mode $\Sigma_c(2P)$ states are also investigated, which are most likely to be narrow states and have good potential to be observed in future experiments.
\end{abstract}

\keywords{$\Lambda_c(2940)$; $^3P_0$ model; Singly heavy baryons}

\maketitle

\section{Introduction}{\label{introduction}}

The singly charmed baryons are composed of one charm quark and two light quarks. Constraints on the nonstrange light quarks, they can be further categorized into the $\Lambda_c$ and $\Sigma_c$ families, which belong to the antisymmetric flavor structure $\bar 3_F$ and symmetric flavor structure $6_F$, respectively. Establishing the spectrum of these charmed baryons has attracted lots of theoretical and experimental attentions~\cite{Cheng:2006dk,Chen:2007xf,Valcarce:2008dr,Ebert:2011kk,Chen:2014nyo,He:2006is,Ortega:2012cx,Ortega:2014eoa,Zhao:2016zhf,Yang:2015eoa,Wang:2018jaj,
Zhong:2007gp,Wang:2017kfr,Liu:2012sj,
Yao:2018jmc,Nagahiro:2016nsx,Dong:2009tg,Chen:2016iyi,Artuso:2000xy,Mizuk:2004yu,Zhu:2000py,Chen:2017sci,Huang:1995ke,Blechman:2003mq,Hwang:2006df,Albertus:2005zy,
Hussain:1999sp,Ivanov:1999bk,Guo:2007qu,Chen:2009tm,Guo:2016wpy,Chen:2016spr,Richard:1992uk,Klempt:2009pi,Crede:2013sze,Cheng:2015iom}. From the Particle Data Group book, there exist nine  $\Lambda_c$ and $\Sigma_c$ baryons, $\Lambda_c(2286)$, $\Lambda_c(2593)$, $\Lambda_c(2625)$, $\Lambda_c(2765)$, $\Lambda_c(2880)$, $\Lambda_c(2940)$, $\Sigma(2455)$, $\Sigma(2520)$, and $\Sigma(2800)$~\cite{Patrignani:2016xqp}. $\Lambda_c(2286)$, $\Sigma(2455)$, and $\Sigma(2520)$ are the $S-$wave ground states, and $\Lambda_c(2593)$ and $\Lambda_c(2625)$ can be well understood as the $P-$wave $\Lambda_c$ states in the conventional quark model. In the $cqq$ configuration, $\Lambda_c(2765)$ and $\Lambda_c(2880)$ might be classified into the $2S$ and $1D$ $\Lambda_c$ states, respectively, while $\Sigma_c(2800)$ is possibly a $1P$ $\Sigma_c$ state. Other conventional or exotic interpretations are also suggested for the $\Lambda_c(2765)$, $\Lambda_c(2880)$, and $\Sigma_c(2800)$ states. Detailed discussions of various assignments and properties can be found in Refs.~\cite{Chen:2016spr,Crede:2013sze,Cheng:2015iom}.

In 2017, the LHCb Collaboration performed an amplitude analysis of the $\Lambda_b^0 \to D^0p\pi^-$ decay process in the $D^0p$ channel, and observed three $\Lambda_c$ resonances, $\Lambda_c(2860)$, $\Lambda_c(2880)$, and $\Lambda_c(2940)$~\cite{Aaij:2017vbw}. Their masses and decay widths were measured as follows,
\begin{eqnarray}
m[\Lambda_c(2860)^+] = 2856.1^{+2.0}_{-1.7}\pm0.5^{+1.1}_{-5.6}~\rm{MeV},
\end{eqnarray}
\begin{eqnarray}
\Gamma[\Lambda_c(2860)^+] = 67.6^{+10.1}_{-8.1}\pm1.4^{+5.9}_{-20.0}~\rm{MeV},
\end{eqnarray}
\begin{eqnarray}
m[\Lambda_c(2880)^+] = 2881.75\pm0.29\pm0.07^{+0.14}_{-0.20}~\rm{MeV},
\end{eqnarray}
\begin{eqnarray}
\Gamma[\Lambda_c(2880)^+] = 5.43^{+0.77}_{-0.71}\pm0.29^{+0.75}_{-0.00}~\rm{MeV},
\end{eqnarray}
\begin{eqnarray}
m[\Lambda_c(2940)^+] = 2944.8^{+3.5}_{-2.5}\pm0.4^{+0.1}_{-4.6}~\rm{MeV},
\end{eqnarray}
\begin{eqnarray}
\Gamma[\Lambda_c(2940)^+] = 27.7^{+8.2}_{-6.0}\pm0.9^{+5.2}_{-10.4}~\rm{MeV}.
\end{eqnarray}
The quantum numbers of $\Lambda_c(2860)$ and $\Lambda_c(2880)$ were determined to be $J^P=\frac{3}{2}^+$ and $J^P=\frac{5}{2}^+$, respectively. The measured information indicates that they may be good candidates of the $1D$-wave $\Lambda_c$ resonances. The spin and parity of the $\Lambda_c(2940)$ state were constrained. The most likely spin-parity quantum numbers of $\Lambda_c(2940)$ are $J^P=\frac{3}{2}^-$, while other possibilities cannot be excluded completely~\cite{Aaij:2017vbw}. With the favorable $J^P=\frac{3}{2}^-$ assignment, the $\Lambda_c(2940)$ may correspond to a conventional $2P$-wave $\Lambda_c$ resonance in the quark model.

In the past years, from the point view of the mass spectrum the properties of $\Lambda_c(2940)$ were attempted to be understood within various quark models. For example, some people studied the $\Lambda_c$ spectrum in the consistent quark model, and found $\Lambda_c(2940)$ could be an excited $\Lambda_c$ state with $J^P=3/2^+$~\cite{Garcilazo:2007eh,Valcarce:2008dr}. Within the diquark picture, $\Lambda_c(2940)$ can be interpreted as the $2P$-wave $\Lambda_c$ resonance with $J^P=1/2^-$ or the $2S$-wave state with $J^P=3/2^+$ in the relativistic quark model~\cite{Ebert:2011kk}, the $2P$-wave $\Lambda_c$ resonance with $J^P=1/2^-$ state in the relativized quark model~\cite{Lu:2016ctt}, and the $J^P=5/2^-$ $1D$-wave state or the $2P$ -wave $\Lambda_c$ resonances in flux tube model~\cite{Chen:2009tm,Chen:2014nyo}. Meanwhile, the $D^*N$ molecular state interpretations were suggested in some works~\cite{He:2006is,Ortega:2012cx,Ortega:2014eoa,Zhao:2016zhf,Yang:2015eoa}, where with the $S-$wave $1/2^-$ or $3/2^-$ assignment, the near threshold behavior of $\Lambda_c(2940)$ can be naturally explained.

Besides the mass spectrum, the $\Lambda_c(2940)$ resonance was also investigated via its decay and production processes.
For example, the strong decays of $\Lambda_c(2940)$ were studied within the chiral perturbation theory,
one found that the spin-parity numbers might be $3/2^+$ or $5/2^-$~\cite{Cheng:2006dk}.
Within the quark model, the strong decays indicated $\Lambda_c(2940)$ can be described as the $D-$wave $\Lambda_c$
state with spin-parity numbers $5/2^+$~\cite{Zhong:2007gp} or $7/2^+$~\cite{Nagahiro:2016nsx}.
Meanwhile, the decay behaviors of the $J^P= 1/2^-$, $3/2^-$, $1/2^+$ $D^*N$ molecule states were investigated~\cite{He:2006is,Dong:2009tg,Ortega:2012cx,Ortega:2014eoa}, and no definitive conclusion was obtained.
Furthermore, the productions of $\Lambda_c(2940)$ in the $\bar p p$, $\pi^-p$, $\gamma n$, and $K^-p$ processes
were studied within effective Lagrangian approaches~\cite{He:2011jp,Dong:2014ksa,Xie:2015zga,Wang:2015rda,Huang:2016ygf}, which provide helpful references for future PANDA and COMPASS experiments.

It is shown that the theoretical works perform lots of interpretations on $\Lambda_c(2940)$, while the quantum numbers $J^P = \frac{3}{2}^-$ determined by LHCb Collaboration favor the conventional $2P$ $\Lambda_c$ resonance or the exotic $D^*N$ molecule description.
Although there are many discussions of $\Lambda_c(2940)$ in the literature as mentioned before,
less discussions of the decay behaviors as the conventional $2P$ $\Lambda_c$ states can be found.
Hence, in this work, we study the strong decays of the $2P$ charmed baryons within the
$^3P_0$ quark pair creation model. Our results indicate that $\Lambda_c(2940)$ as the $\lambda-$mode $\Lambda_{c1}(\frac{1}{2}^-,2P)$ and $\Lambda_{c1}(\frac{3}{2}^-,2P)$ states are both allowed, and the $J^P=3/2^-$ state
$\Lambda_{c1}(\frac{3}{2}^-,2P)$ is more favorable.

This paper is organized as follows. The $^3P_0$ model is briefly introduced in Sec.~\ref{model}. The strong decays of the $2P$ $\Lambda_c$ and $\Sigma_c$ charmed baryons are estimated in Sec.~\ref{decay}. A short summary is presented in the last section.

\section{$^3P_0$ Model}{\label{model}}

In this work, we adopt the $^3P_0$ model to calculate the
Okubo-Zweig-Iizuka-allowed two-body strong decays of the $2P$ $\Lambda_c$ and $\Sigma_c$ states. The $^3P_0$ model, also known as the quark pair creation model, has been extensively employed to study the strong decays with considerable successes~\cite{3p0model1,3p0model2,3p0model3,3p0model4,3p0model5,3p0model6,rss,micu,Chen:2007xf,Zhao:2016qmh,Ye:2017yvl,Ye:2017dra,Zhao:2017fov,Chen:2017gnu,Chen:2016iyi,
Lu:2014zua,Lu:2016bbk,Ferretti:2014xqa,Godfrey:2015dva,Segovia:2012cd}. In this model, the hadrons decay occurs through a quark-antiquark pair with the vacuum quantum number $J^{PC}=0^{++}$~\cite{micu}. Here we perform a brief review of the $^3P_0$ model. In the nonrelativistic limit, the transition operator $T$ of the decay $A\rightarrow BC$ in the
$^3P_0$ model can be assumed as~\cite{Chen:2007xf,Chen:2017gnu}
\begin{eqnarray}
T&=&-3\gamma\sum_m\langle 1m1-m|00\rangle\int
d^3\boldsymbol{p}_4d^3\boldsymbol{p}_5\delta^3(\boldsymbol{p}_4+\boldsymbol{p}_5)\nonumber\\
&&\times {\cal{Y}}^m_1\left(\frac{\boldsymbol{p}_4-\boldsymbol{p}_5}{2}\right
)\chi^{45}_{1,-m}\phi^{45}_0\omega^{45}_0b^\dagger_{4i}(\boldsymbol{p}_4)d^\dagger_{4j}(\boldsymbol{p}_5),
\end{eqnarray}
where $\gamma$ is a dimensionless $q_4\bar{q}_5$ pair-production strength, and $\boldsymbol{p}_4$ and
$\boldsymbol{p}_5$ are the momenta of the created quark $q_4$ and
antiquark  $\bar{q}_5$, respectively. The $i$ and $j$ are the color indices of the created quark and antiquark. $\phi^{45}_{0}=(u\bar u + d\bar d +s\bar s)/\sqrt{3}$,
$\omega^{45}=\delta_{ij}$, and $\chi_{{1,-m}}^{45}$ are the flavor singlet, color singlet,
and spin triplet wave functions of the  $q_4\bar{q}_5$, respectively. The
solid harmonic polynomial
${\cal{Y}}^m_1(\boldsymbol{p})\equiv|p|Y^m_1(\theta_p, \phi_p)$ reflects
the $P-$wave momentum-space distribution of the $q_4\bar{q}_5$ quark pair.

For the initial baryon $A$, we adopt the definition of the mock states~\cite{Hayne:1981zy}
\begin{eqnarray}
&&|A(n^{2S_A+1}_AL_{A}\,\mbox{}_{J_A M_{J_A}})(\boldsymbol{P}_A)\rangle
\equiv \nonumber\\
&& \sqrt{2E_A}\sum_{M_{L_A},M_{S_A}}\langle L_A M_{L_A} S_A
M_{S_A}|J_A
M_{J_A}\rangle \int d^3\boldsymbol{p}_1d^3\boldsymbol{p}_2d^3\boldsymbol{p}_3\nonumber\\
&&\times \delta^3(\boldsymbol{p}_1+\boldsymbol{p}_2+\boldsymbol{p}_3-\boldsymbol{P}_A)\psi_{n_AL_AM_{L_A}}(\boldsymbol{p}_1,\boldsymbol{p}_2,\boldsymbol{p}_3)\chi^{123}_{S_AM_{S_A}}
\phi^{123}_A\omega^{123}_A\nonumber\\
&&\times  \left|q_1(\boldsymbol{p}_1)q_2(\boldsymbol{p}_2)q_3(\boldsymbol{p}_3)\right\rangle,
\end{eqnarray}
which satisfies the normalization condition
\begin{eqnarray}
\langle A(\boldsymbol{P}_A)|A(\boldsymbol{P}^\prime_A)\rangle=2E_A\delta^3(\boldsymbol{P}_A-\boldsymbol{P}^\prime_A).
\end{eqnarray}
The $\boldsymbol{p}_1$, $\boldsymbol{p}_2$, and $\boldsymbol{p}_3$ are the momenta of the
quarks $q_1$, $q_2$, and $q_3$, respectively. $\boldsymbol{P}_A$ denotes the momentum of the initial state $A$.
$\chi^{123}_{S_AM_{S_A}}$, $\phi^{123}_A$, $\omega^{123}_A$,
$\psi_{n_AL_AM_{L_A}}(\boldsymbol{p}_1,\boldsymbol{p}_2,\boldsymbol{p}_3)$ are the spin, flavor, color, and
space wave functions of the baryon $A$ composed of $q_1q_2q_3$ with total energy $E_A$, respectively. The definitions of the mock states $B$ and $C$ are similar to that of initial state $A$, and can be find in Ref.~\cite{Chen:2007xf}.

\begin{figure}[!htbp]
\includegraphics[scale=0.85]{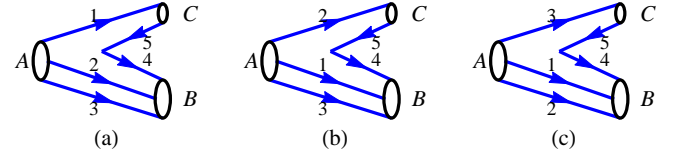}
\vspace{0.0cm} \caption{The baryon decay process $A\to B+C$ in the $^3P_0$ model.}
\label{qpc}
\end{figure}

For the decay of the charmed baryon $A$, three possible rearrangements exist,
\begin{eqnarray}
A(q_1,q_2,c_3)+P(q_4,\bar q_5)\to B(q_2,q_4,c_3)+C(q_1,\bar q_5),\\
A(q_1,q_2,c_3)+P(q_4,\bar q_5)\to B(q_1,q_4,c_3)+C(q_2,\bar q_5),\\
A(q_1,q_2,c_3)+P(q_4,\bar q_5)\to B(q_1,q_2,q_4)+C(c_3,\bar q_5),
\end{eqnarray}
where the $q_i$ and $c_3$ denote the light quark and charm quark, respectively. These three ways of recouplings are also shown in Figure~\ref{qpc}.

The $S$ matrix can be defined as
\begin{eqnarray}
\langle
f|S|i\rangle=I-i2\pi\delta(E_f-E_i){\cal{M}}^{M_{J_A}M_{J_B}M_{J_C}},
\end{eqnarray}
where the ${\cal{M}}^{M_{J_A}M_{J_B}M_{J_C}}$ is the helicity amplitude of the decay process $A\to B+C$. Taken the process $A(q_1,q_2,c_3)+P(q_4,\bar q_5)\to B(q_1,q_4,c_3)+C(q_2,\bar q_5)$ shown in Fig. 1(b) as an example, the helicity amplitude ${\cal{M}}^{M_{J_A}M_{J_B}M_{J_C}}$ reads~\cite{Chen:2007xf,Ye:2017yvl,Ye:2017dra},
\begin{eqnarray}
&&\delta^3(\boldsymbol{p}_B+\boldsymbol{p}_C-\boldsymbol{p}_A){\cal{M}}^{M_{J_A}M_{J_B}M_{J_C}} = \nonumber\\
&&- \gamma \sqrt{8E_AE_BE_C} \sum_{M_{\rho_A}} \sum_{M_{L_A}} \sum_{M_{\rho_B}} \sum_{M_{L_B}} \sum_{M_{S_1}, M_{S_3}, M_{S_4}, m}\nonumber\\
&& \times  \langle J_{l_A} M_{J_{l_A}}S_3M_{S_3}|J_AM_{J_A}\rangle \langle L_{\rho_A} M_{L_{\rho_A}}L_{\Lambda_A}M_{L_{\Lambda_A}}|L_AM_{L_A} \rangle \nonumber\\ && \times \langle L_A M_{L_A}S_{12}M_{S_{12}}|J_{l_A}M_{J_{l_A}}\rangle \langle S_1M_{S_1}S_2M_{S_2}|S_{12}M_{S_{12}} \rangle \nonumber\\
&& \times \langle J_{l_B} M_{J_{l_B}}S_3M_{S_3}|J_BM_{J_B}\rangle \langle L_{\rho_B} M_{L_{\rho_B}}L_{\Lambda_B}M_{L_{\Lambda_B}}|L_BM_{L_B}\rangle \nonumber\\ && \times \langle L_B M_{L_B}S_{14}M_{S_{14}}|J_{l_B}M_{J_{l_B}}\rangle \langle S_1M_{S_1}S_4M_{S_4}|S_{14}M_{S_{14}}\nonumber\\
&& \times \langle 1m 1-m|00\rangle \langle S_4M_{S_4}S_5M_{S_5}|1-m \rangle \nonumber\\
&& \times \langle L_C M_{L_C}S_CM_{S_C}|J_CM_{J_C}\rangle \langle S_2M_{S_2}S_5M_{S_5}|S_CM_{S_C}\rangle \nonumber\\
&& \times \langle \phi_B^{143} \phi_C^{25}|\phi_A^{123}\phi_0^{45}\rangle I^{M_{L_A}m}_{M_{L_B}M_{L_C}}(\boldsymbol{p}),
\end{eqnarray}
where $\langle \phi_B^{143} \phi_C^{25}|\phi_A^{123}\phi_0^{45}\rangle$ are the overlap of the flavor wavefunctions. The $I^{M_{L_A}m}_{M_{L_B}M_{L_C}}(\boldsymbol{p})$ are the spatial overlaps of the initial and final states, which can be written as
\begin{eqnarray}
I^{M_{L_A}m}_{M_{L_B}M_{L_C}}(\boldsymbol{p}) & = & \int d^3\boldsymbol{p}_1d^3\boldsymbol{p}_2d^3\boldsymbol{p}_3d^3\boldsymbol{p}_4d^3\boldsymbol{p}_5  \nonumber\\ && \times \delta^3(\boldsymbol{p}_1+\boldsymbol{p}_2+\boldsymbol{p}_3-\boldsymbol{P}_A)\delta^3(\boldsymbol{p}_4+\boldsymbol{p}_5)\nonumber\\ && \times \delta^3(\boldsymbol{p}_1+\boldsymbol{p}_4+\boldsymbol{p}_3-\boldsymbol{P}_B)\delta^3(\boldsymbol{p}_2+\boldsymbol{p}_5-\boldsymbol{P}_C) \nonumber\\
&& \times \psi^*_B(\boldsymbol{p}_1,\boldsymbol{p}_4,\boldsymbol{p}_3) \psi^*_C(\boldsymbol{p}_2,\boldsymbol{p}_5)\nonumber\\
&& \times\psi_A(\boldsymbol{p}_1,\boldsymbol{p}_2,\boldsymbol{p}_3){\cal{Y}}^m_1\left(\frac{\boldsymbol{p}_4-\boldsymbol{p}_5}{2}\right
).
\end{eqnarray}

In this issue, we employ the simplest vertex which assumes a
spatially constant pair production strength $\gamma$~\cite{micu}, the relativistic phase space, and the simple harmonic oscillator wave functions. With the relativistic phase space, the decay width
$\Gamma(A\rightarrow BC)$ can be expressed as follows
\begin{eqnarray}
\Gamma= \pi^2\frac{p}{M^2_A}\frac{s}{2J_A+1}\sum_{M_{J_A},M_{J_B},M_{J_C}}|{\cal{M}}^{M_{J_A}M_{J_B}M_{J_C}}|^2,
\end{eqnarray}
where $p=|\boldsymbol{p}|=\frac{\sqrt{[M^2_A-(M_B+M_C)^2][M^2_A-(M_B-M_C)^2]}}{2M_A}$,
and $M_A$, $M_B$, and $M_C$ are the masses of the hadrons $A$, $B$,
and $C$, respectively. $s=1/(1+\delta_{BC})$ is a statistical factor which is needed if $B$ and $C$ are identical particles. Due to $B$ and $C$ correspond to baryon and meson, respectively, the $s$ always equals to one in this work.

\section{Strong decay}{\label{decay}}

\subsection{Notations and parameters}
In our calculation, we adopt the same notations of $\Lambda_c$, $\Sigma_c$ and $\Xi_c$ baryons as those in Ref.~\cite{Chen:2016spr,Chen:2007xf}. For the spatial $2P$ excited states, the symbol $2P$ are added. In Table~\ref{tab1}, The $n_\rho$ and $L_\rho$ stand the nodal and orbital angular momentum between the two light quarks, while $n_\lambda$ and $L_\lambda$ denote the nodal and angular momentum between the two light quark system and the charm quark. $L$ is the total orbital angular momentum, $S_\rho$ is the total spin of the two light quarks, $J_l$ is total angular momentum of $L$ and $S_\rho$, and $J$ is the total angular momentum.

\begin{table}[!htbp]
\begin{center}
\caption{ \label{tab1} Notations and quantum numbers of the relevant $\Lambda_c$, $\Sigma_c$ and $\Xi_c$ baryons.}
\normalsize
\renewcommand{\arraystretch}{1.5}
\begin{tabular}{p{1.5cm}<{\centering}p{0.7cm}<{\centering}p{0.7cm}<{\centering}p{0.7cm}<{\centering}p{0.7cm}<{\centering}p{0.7cm}<{\centering}
p{0.7cm}<{\centering}p{0.7cm}<{\centering}p{0.7cm}<{\centering}}
\hline\hline
 State                          & $n_\rho$ & $L_\rho$     &  $n_\lambda$      &  $L_\lambda$  &  $L$  & $S_\rho$ & $J_l$  & $J$           \\\hline
 $\Lambda_c$     & 0        & 0            &  0                &  0            &  0    & 0        & 0      & $\frac{1}{2}$ \\
 $\Sigma_c$      & 0        & 0            &  0                &  0            &  0    & 1        & 1      & $\frac{1}{2}$ \\
 $\Sigma_c^*$    & 0        & 0            &  0                &  0            &  0    & 1        & 1      & $\frac{3}{2}$ \\
 $\Xi_c$     & 0        & 0            &  0                &  0            &  0    & 0        & 0      & $\frac{1}{2}$ \\
 $\Xi_c^\prime$      & 0        & 0            &  0                &  0            &  0    & 1        & 1      & $\frac{1}{2}$ \\
 $\Xi_c^{\prime *}$    & 0        & 0            &  0                &  0            &  0    & 1        & 1      & $\frac{3}{2}$ \\
 $\Lambda_{c1}(\frac{1}{2}^-,2P)$     & 0        & 0            &  1                &  1            &  1    & 0        & 1      & $\frac{1}{2}$ \\
 $\Lambda_{c1}(\frac{3}{2}^-,2P)$     & 0        & 0            &  1                &  1            &  1    & 0        & 1      & $\frac{3}{2}$ \\
 $\Sigma_{c0}(\frac{1}{2}^-,2P)$     & 0        & 0            &  1                &  1            &  1    & 1        & 0      & $\frac{1}{2}$ \\
 $\Sigma_{c1}(\frac{1}{2}^-,2P)$     & 0        & 0            &  1                &  1            &  1    & 1        & 1      & $\frac{1}{2}$ \\
 $\Sigma_{c1}(\frac{3}{2}^-,2P)$     & 0        & 0            &  1                &  1            &  1    & 1        & 1      & $\frac{3}{2}$ \\
 $\Sigma_{c2}(\frac{3}{2}^-,2P)$     & 0        & 0            &  1                &  1            &  1    & 1        & 2      & $\frac{3}{2}$ \\
 $\Sigma_{c2}(\frac{5}{2}^-,2P)$     & 0        & 0            &  1                &  1            &  1    & 1        & 2      & $\frac{5}{2}$ \\
 $\tilde \Lambda_{c0}(\frac{1}{2}^-,2P)$     & 1        & 1            &  0                &  0            &  1    & 1        & 0      & $\frac{1}{2}$ \\
 $\tilde \Lambda_{c1}(\frac{1}{2}^-,2P)$     & 1        & 1            &  0                &  0            &  1    & 1        & 1      & $\frac{1}{2}$ \\
 $\tilde \Lambda_{c1}(\frac{3}{2}^-,2P)$     & 1        & 1            &  0                &  0            &  1    & 1        & 1      & $\frac{3}{2}$ \\
 $\tilde \Lambda_{c2}(\frac{3}{2}^-,2P)$     & 1        & 1            &  0                &  0            &  1    & 1        & 2      & $\frac{3}{2}$ \\
 $\tilde \Lambda_{c2}(\frac{5}{2}^-,2P)$     & 1        & 1            &  0                &  0            &  1    & 1        & 2      & $\frac{5}{2}$ \\
 $\tilde \Sigma_{c1}(\frac{1}{2}^-,2P)$      & 1        & 1            &  0                &  0            &  1    & 0        & 1      & $\frac{1}{2}$ \\
 $\tilde \Sigma_{c1}(\frac{3}{2}^-,2P)$      & 1        & 1            &  0                &  0            &  1    & 0        & 1      & $\frac{3}{2}$ \\
 \hline\hline
\end{tabular}
\end{center}
\end{table}
For the masses of the two $\Lambda_{c1}(2P)$ states, we adopt the mass of $\Lambda(2940)$ from LHCb experimental data. Masses of the other $2P$ states are taken from theoretical predictions. For the final ground states, their masses are adopted from the Particle Data Group~\cite{Patrignani:2016xqp}. For the harmonic oscillator parameters of mesons, we use the effective values obtained by relativized quark model, i.e., $R= 2.5~\rm{GeV^{-1}}$ for $\pi/\rho/\omega/K/\eta$ meson, $R= 1.67~\rm{GeV^{-1}}$ for $D$ meson, $R= 1.94~\rm{GeV^{-1}}$ for $D^*$ meson, and $R= 1.54~\rm{GeV^{-1}}$ for $D_s$ meson~\cite{Godfrey:2015dva}. For the baryon parameters, we use $\alpha_\rho=400~\rm{MeV}$ and
\begin{eqnarray}
\alpha_\lambda=\Bigg(\frac{3m_Q}{2m_q+m_Q} \Bigg)^\frac{1}{4} \alpha_\rho,
\end{eqnarray}
where the $m_Q$ and $m_q$ are the heavy and light quark masses, respectively~\cite{Zhong:2007gp}. The $m_{u/d}=220~\rm{MeV}$, $m_s=419~\rm{MeV}$, and $m_c=1628~\rm{MeV}$ are introduced to explicitly break the SU(4) symmetry~\cite{Godfrey:1985xj,Capstick:1986bm,Godfrey:2015dva}.
There is an overall parameter $\gamma$, which is determined by the well determined width of the $\Sigma_c(2520)^{++} \to \Lambda_c \pi^+$
process. The $\gamma=9.83$ is obtained by reproducing the width, $\Gamma [\Sigma_c(2520)^{++} \to \Lambda_c \pi^+]=14.78~\rm{MeV}$~\cite{Patrignani:2016xqp}.

\begin{figure}[!htbp]
\includegraphics[scale=0.70]{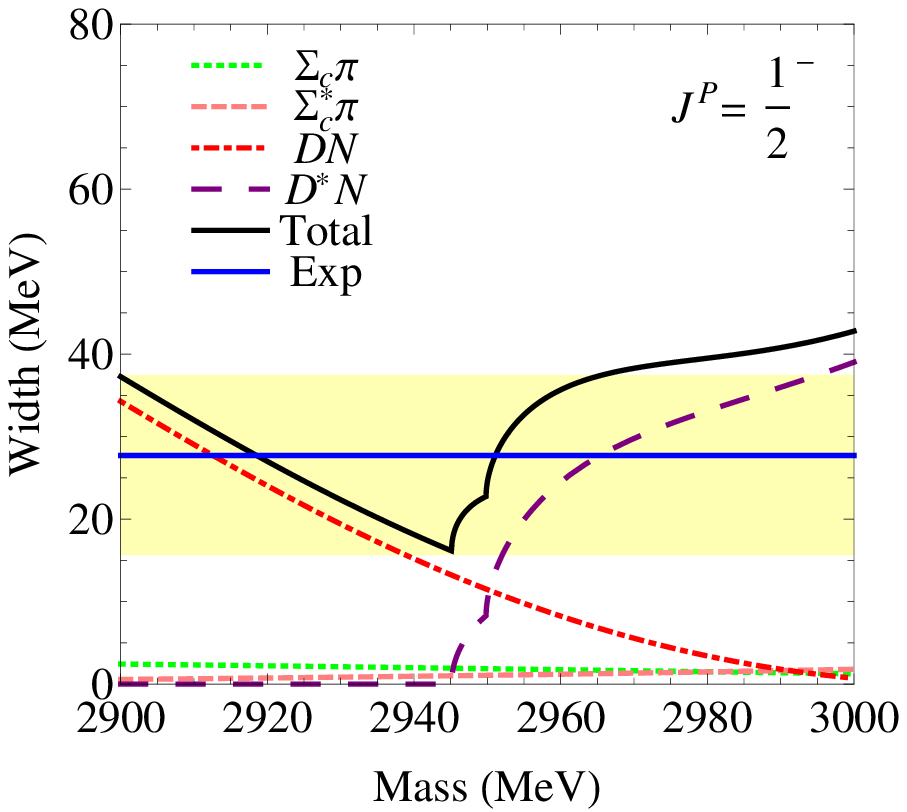}
\includegraphics[scale=0.70]{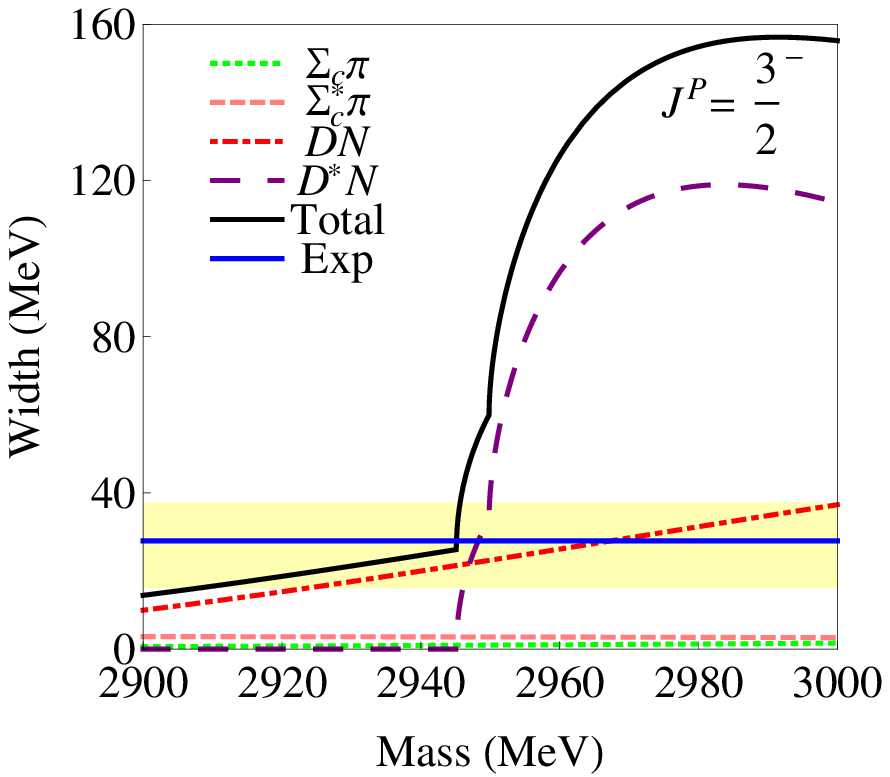}
\vspace{0.0cm} \caption{The decay widths of the $\Lambda_{c1}(\frac{1}{2}^-,2P)$ and $\Lambda_{c1}(\frac{3}{2}^-,2P)$ states as functions of the initial state mass. The blue line with a yellow band denotes the LHCb experimental data.}
\label{2940}
\end{figure}

\begin{figure}[!htbp]
\includegraphics[scale=0.70]{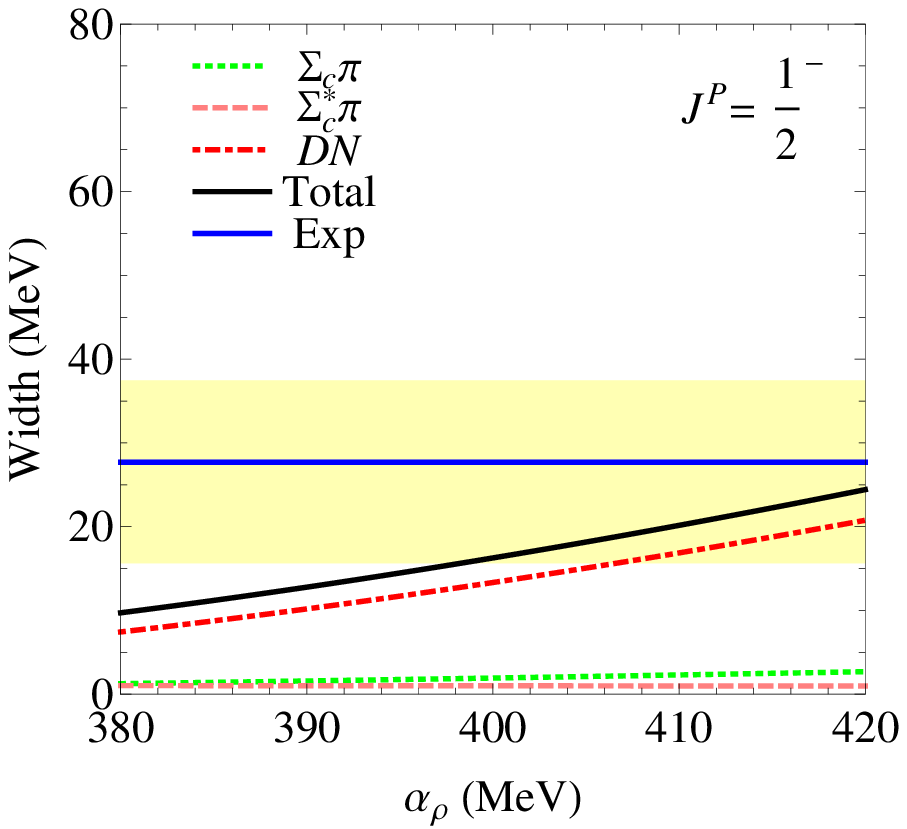}
\includegraphics[scale=0.70]{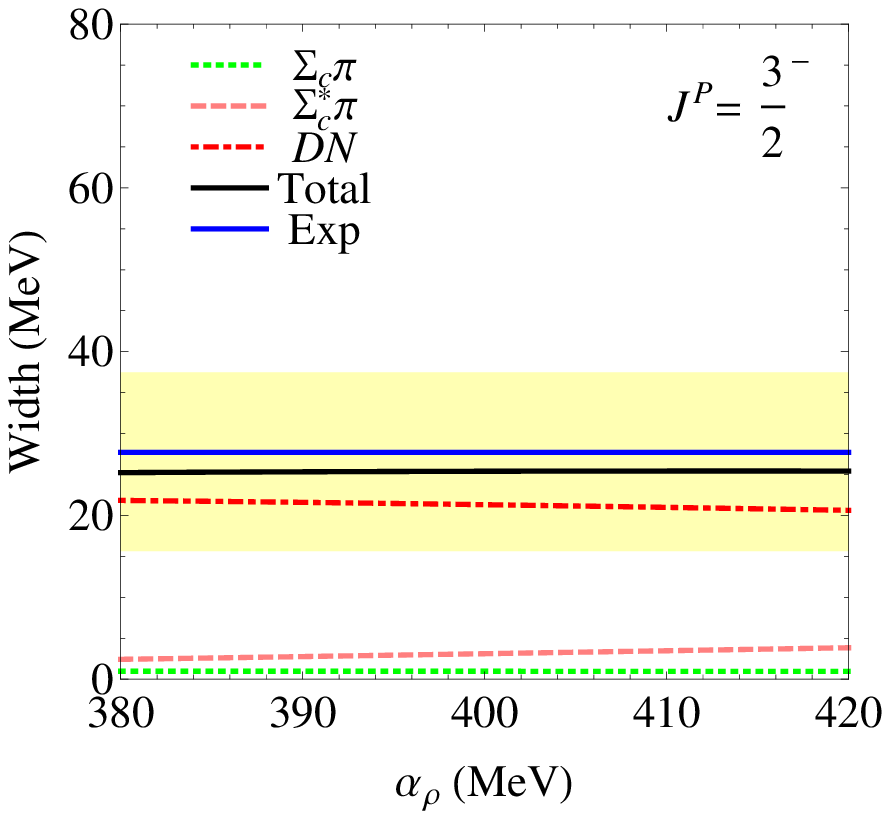}
\vspace{0.0cm} \caption{The dependence of the decay widths on the harmonic oscillator parameters $\alpha_\rho$.}
\label{par}
\end{figure}

\subsection{$\Lambda_c(2940)$}

The strong decays of $\Lambda_c(2940)$ as the $\Lambda_{c1}(\frac{1}{2}^-,2P)$ and $\Lambda_{c1}(\frac{3}{2}^-,2P)$ assignments are calculated. The results are listed in Table~\ref{tab2}. It is shown that the total decay widths of the $\Lambda_{c1}(\frac{1}{2}^-,2P)$ and $\Lambda_{c1}(\frac{3}{2}^-,2P)$ states are 16.27 MeV and 25.39 MeV, respectively. Compared with the experimental total width $27.7^{+8.2}_{-6.0}\pm0.9^{+5.2}_{-10.4}~\rm{MeV}$ measured by LHCb Collaboration, both assignments are allowed. However, the $J^P=\frac{3}{2}^-$ assignment is more favorable. The main decay mode is the $DN$ channel, and the partial decay widths of the $\Sigma_c \pi$ and $\Sigma^*_c \pi$ channels are rather small, which is consistent with the fact that $\Lambda_c(2940)$ was observed in $D^0p$ invariant mass distribution. The partial decay width ratios of the $J^P=\frac{1}{2}^-$ state are predicted to be
\begin{eqnarray}
\Gamma[\Sigma_c \pi]:\Gamma[\Sigma_c^* \pi]:\Gamma[DN]  = 1:0.52:6.91,
\end{eqnarray}
and the partial decay width ratios of the $J^P=\frac{3}{2}^-$ state are predicted to be
\begin{eqnarray}
\Gamma[\Sigma_c \pi]:\Gamma[\Sigma_c^* \pi]:\Gamma[DN] = 1:3.22:22.03.
\end{eqnarray}
These ratios are independent with the overall parameter $\gamma$ in the $^3P_0$ model, and the divergence of these two set of quantum number assignments can be tested in future experimental data.

\begin{table}
\begin{center}
\caption{ \label{tab2} Decay widths of the $\Lambda_c(2940)$ as the $\Lambda_{c1}(\frac{1}{2}^-,2P)$ and
$\Lambda_{c1}(\frac{3}{2}^-,2P)$ in MeV.}
\renewcommand{\arraystretch}{1.5}
\normalsize
\begin{tabular}{lcccc}
\hline\hline
   Mode                            &\multicolumn{2}{c} {$\Lambda_c(2940)$}\\
                               & $\Lambda_{c1}(\frac{1}{2}^-,2P)$      & $\Lambda_{c1}(\frac{3}{2}^-,2P)$     \\\hline
  $\Sigma_c^{++} \pi^-$        & 0.65            & 0.32             \\
  $\Sigma_c^{+} \pi^0$        & 0.64            & 0.33             \\
  $\Sigma_c^{0} \pi^+$        & 0.65            & 0.32             \\
  $\Sigma_c^{*++} \pi^-$        & 0.33            & 1.04             \\
  $\Sigma_c^{*+} \pi^0$        & 0.34            & 1.03             \\
  $\Sigma_c^{*0} \pi^+$        & 0.33            & 1.04             \\
  $D^+n$                       & 7.21            & 10.26          \\
  $D^0p$                       & 6.13            & 11.05              \\
  Total width                  & 16.27           & 25.39       \\
  Experiment                   &\multicolumn{2}{c} {$27.7^{+8.2}_{-6.0}\pm0.9^{+5.2}_{-10.4}$}\\
\hline\hline
\end{tabular}
\end{center}
\end{table}

In Figure.~\ref{2940}, we plot the variation of the decay widths as a function of the initial baryon mass. It is seen that the partial width of the $DN$ channel decreases for the $1/2^-$ state, while increases for the $3/2^-$ state. The $\Sigma_c \pi$ and $\Sigma_c^* \pi$ decay modes are small enough in this mass region. When the mass lies above the $D^*N$ threshold, the $D^*N$ channel also performs significant contributions to the total decay widths in both cases. Since the mass splitting of $\Lambda_c(1P)$ is
\begin{eqnarray}
m[\Lambda_c(2625)]-m[\Lambda_c(2595)]  = 36~\rm{MeV},
\end{eqnarray}
the mass splitting of the two $\Lambda_c(2P)$ states is smaller than $36~\rm{MeV}$. Considering $\Lambda_c(2940)$ as the $\Lambda_{c1}(\frac{3}{2}^-,2P)$ state, the mass of the $\Lambda_{c1}(\frac{1}{2}^-,2P)$ state should lie in $2909\sim 2945 ~\rm{MeV}$. From Fig.~\ref{2940}, the $\Lambda_{c1}(\frac{1}{2}^-,2P)$ state has a width of $16\sim 33 ~\rm{MeV}$, which can be searched in the $DN$ final state in future experiments.

The dependence on the harmonic oscillator parameter $\alpha_\rho$ is also investigated in Fig.~\ref{par}.  When the $\alpha_\rho$ increases, the total decay width also increases for the $1/2^-$ state. While, the total decay width of the $3/2^-$ state is almost unchanged with the $\alpha_\rho$ variation. Within this reasonable range of the parameter $\alpha_\rho$, our conclusions remain.

\begin{table}
\begin{center}
\caption{ \label{tab3} Predicted masses for the $\lambda$-mode $\Sigma_c(2P)$ states in the literature. The units are in MeV.}
\renewcommand{\arraystretch}{1.5}
\small
\begin{tabular}{p{1.2cm}<{\centering}p{1.3cm}<{\centering}p{1.3cm}<{\centering}p{1.3cm}<{\centering}p{1.3cm}<{\centering}p{1.3cm}<{\centering}}
\hline\hline
 State        & RQM\cite{Capstick:1986bm}& RQM\cite{Ebert:2007nw} &  RQM\cite{Ebert:2011kk}  &  HCQM\cite{Shah:2016mig}&  NRQM\cite{Chen:2016iyi} \\
 $\Sigma_{c0}(\frac{1}{2}^-,2P)$     & 3185        & 3186            &  3172                &  3245            &  2971   \\
 $\Sigma_{c1}(\frac{1}{2}^-,2P)$     & 3195        & 3176            &  3125                &  3256            &  3018    \\
 $\Sigma_{c1}(\frac{3}{2}^-,2P)$     & 3195        & 3180            &  3172                &  3223            &  3036    \\
 $\Sigma_{c2}(\frac{3}{2}^-,2P)$     & 3210        & 3147            &  3151                &  3233            &  3044    \\
 $\Sigma_{c2}(\frac{5}{2}^-,2P)$     & 3220        & 3167            &  3161                &  3203            &  3040   \\
\hline\hline
\end{tabular}
\end{center}
\end{table}

\begin{table}
\begin{center}
\caption{ \label{sig} The strong decay behaviors of the five $\lambda$-mode $\Sigma_c(2P)$ states. The masses of the initial baryons are taken from Ref.~\cite{Capstick:1986bm}.  The units are in MeV.}
\renewcommand{\arraystretch}{1.5}
\small
\begin{tabular}{p{0.65cm}<{\centering}p{1.45cm}<{\centering}p{1.45cm}<{\centering}p{1.45cm}<{\centering}p{1.45cm}<{\centering}p{1.45cm}<{\centering}}
\hline\hline
 Mode     & $\Sigma_{c0}(\frac{1}{2}^-,2P)$  & $\Sigma_{c1}(\frac{1}{2}^-,2P)$  & $\Sigma_{c1}(\frac{3}{2}^-,2P)$ &  $\Sigma_{c2}(\frac{3}{2}^-,2P)$  &  $\Sigma_{c2}(\frac{5}{2}^-,2P)$  \\
 $\Lambda_c \pi^+$          & 1.04       & $-$             &  $-$                 &  2.20            &  2.20   \\
 $\Lambda_c \rho^+$         & $-$        & 3.09            &  3.09                &  0.92            &  1.05    \\
 $\Sigma_c^{++} \pi^0$      & $-$        & 0.01            &  0.69                &  1.29            &  0.59    \\
 $\Sigma_c^{+} \pi^+$       & $-$        & 0.01            &  0.69                &  1.29            &  0.59    \\
 $\Sigma_c^{*++} \pi^0$     & $-$        & 1.09            &  0.60                &  1.05            &  1.69   \\
 $\Sigma_c^{*+} \pi^+$      & $-$        & 1.09            &  0.61                &  1.05            &  1.69  \\
 $\Sigma_c^{++} \eta$       & $-$        & 0.86            &  0.12                &  0.26            &  0.13    \\
 $\Sigma_c^{*++} \eta$      & $-$        & 0.10            &  1.51                &  0.12            &  0.21    \\
 $\Xi_c^{+} K^+$            & 4.93       & $-$             &  $-$                 &  0.96            &  1.04    \\
 $\Xi_c^{\prime +} K^+$     & $-$        & 5.13            &  0.08                &  0.19            &  0.10    \\
 $\Xi_c^{\prime *+} K^+$    & $-$        & 0.02            &  5.01                &  0.03            &  0.08    \\
 $D^+ p$                    & 4.17       & 8.98            &  3.21                &  0.62            &  9.71    \\
 $D^{*+} p$                 & 1.08      & 11.63            &  8.63                &  34.46            &  21.77    \\
 $D^0 \Delta^{++}$          & 9.44       & 5.82            &  16.16               &  18.16            &  12.33    \\
 $D^+ \Delta^{+}$           & 2.83       & 1.77            &  5.22                &  6.70           &  3.85    \\
 $D_s \Sigma^{+}$           & 4.56       & 8.92            &  0.07                &  0.03            &  0.70    \\
 Total                      & 28.06      & 48.52           &  45.68               &  69.32            &  57.73    \\
\hline\hline
\end{tabular}
\end{center}
\end{table}

\begin{figure}[!htbp]
\includegraphics[scale=0.32]{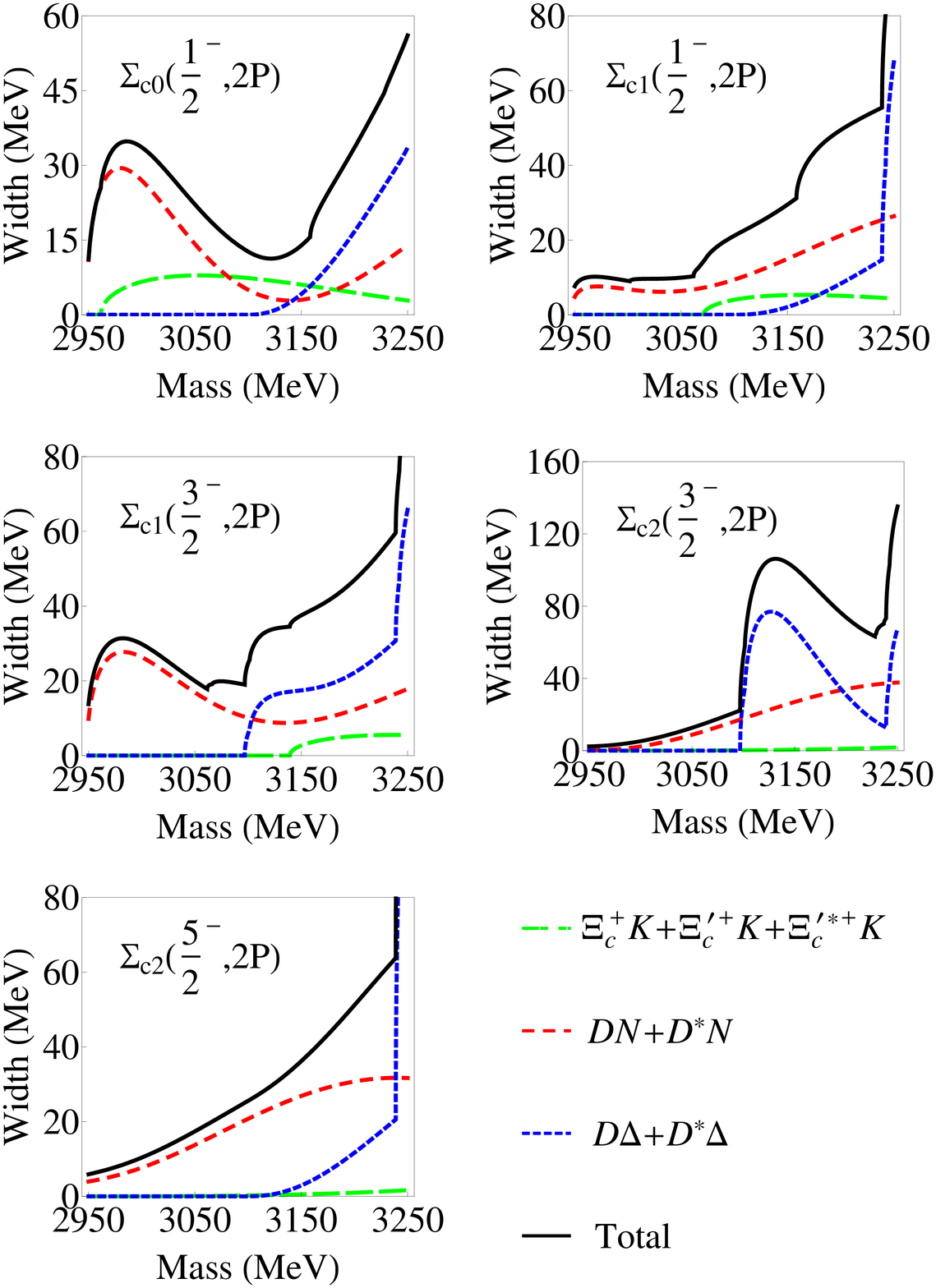}
\vspace{0.0cm} \caption{The decay widths of the five $\lambda$-mode $\Sigma_c(2P)$ states as functions of their masses. The partial decay widths of $\Lambda_c \pi$, $\Lambda_c \rho$, $\Sigma_c \pi$, $\Sigma_c^* \pi$, $\Sigma_c \eta$, $\Sigma_c^*\eta$, $\Sigma_c \rho$, $\Sigma_c \omega$, and $D_s \Sigma$ channels are relatively small, which are not presented here.}
\label{sigmac}
\end{figure}

\subsection{$\Sigma_c(2P)$}

There are five $\lambda$-mode $\Sigma_c(2P)$ states, denoted as $\Sigma_{c0}(\frac{1}{2}^-,2P)$, $\Sigma_{c1}(\frac{1}{2}^-,2P)$, $\Sigma_{c1}(\frac{3}{2}^-,2P)$, $\Sigma_{c2}(\frac{3}{2}^-,2P)$, and $\Sigma_{c2}(\frac{5}{2}^-,2P)$, respectively. Although no information exists for these states in the experiments, some theoretical works have investigated their masses~\cite{Capstick:1986bm,Ebert:2007nw,Ebert:2011kk,Shah:2016mig,Chen:2016iyi}. In Tab.~\ref{tab3}, we collect the predicted masses of $\lambda$-mode $\Sigma_c(2P)$ states in the literature. Here, we employ the masses predicted by the relativized quark model~\cite{Capstick:1986bm} to calculate their strong decays, and the results are listed in Tab.~\ref{sig}. The total decay widths of these five states are about $28\sim69~\rm{MeV}$, which are relatively narrow. The main decay modes are light baryon plus heavy meson channels, while the heavy baryon plus light meson channels are rather small. The narrow total decay widths and large $D^{(*)}N$ branching ratios suggest that these states have good potential to be observed in future experiments. Moreover, the decay widths as functions of their initial masses are plotted in Fig.~\ref{sigmac} for reference.

There are also $\rho$-mode excited $2P$ states, where a symbol $``\sim "$ are added to distinguish them from the $\lambda$-mode states in Tab~\ref{tab1}. The theoretical predictions of these states are scarce. In the singly heavy baryon sector, exciting the $\lambda$-mode is much easier than the $\rho$-mode, hence, the $\rho$-mode excited $2P$ states should be much higher than the $\lambda$-mode states. With the higher masses, more strong decay channels will be open. Due to the lack of mass information and the uncertainties of many decay channels, it seems untimely to study their properties in present work.

\section{Summary}{\label{Summary}}

 In this work, we study the strong decays of the $\Lambda_c(2940)$ baryon within the $^3P_0$ model. Considering the mass, parity and $D^0 p$ decay mode, we tentatively assign $\Lambda_c(2940)$ as the $\lambda$-mode $\Lambda_c(2P)$ states. The main decay mode is $DN$ channel for both $1/2^-$ and $3/2^-$ states. The total decay width of the $\Lambda_{c1}(\frac{1}{2}^-,2P)$ and $\Lambda_{c1}(\frac{3}{2}^-,2P)$ states are 16.27 MeV and 25.39 MeV, respectively. Compared with the total width measured by LHCb Collaboration, both assignments are allowed, and the $J^P=\frac{3}{2}^-$ assignment is more favorable. Other $\lambda-$mode $\Sigma_c(2P)$ states are also investigated. The relatively narrow total decay widths and large $D^{(*)}N$ branching ratios can be tested in future experimental searches.

\bigskip
\noindent
\begin{center}
{\bf ACKNOWLEDGEMENTS}\\

\end{center}
We would like to thank Yu-Bing Dong, De-Min Li and Yin-Huang for valuable discussions. This project is supported by
the National Natural Science Foundation of China under Grants No. 11705056 and No. 11775078. This work is also in part supported by China Postdoctoral
Science Foundation under Grant No.~2017M620492.

\end{document}